# Observation of the toroidal rotation in a new designed compact torus system for EAST

Z. H. Zhao[1,2], T. Lan[4,*], D. F. Kong[2,3,], Y. Ye[2], S. B. Zhang[3], G. Zhuang, X. H. Zhang[1,7,*], G. H. Hu[3], C. Chen[4], J. Wu[4], S. Zhang[4], M. B. Qi[1], C. H. Li[1], X. M. Yang[1], L. Y. Nie[1], F. Wen[4], P. F. Zi[4], L. Li[4], F. W. Meng[1,2], B. Li[1,2], Q. L. Dong[2,5], Y. Q. Huang[2,6]

[1] School of Computer Science and Information Engineering, Hefei University of Technology, Hefei, People's Republic of China
[2] Institute of Energy, Hefei Comprehensive National Science Center, Hefei, People's Republic of China
[3] Institute of Plasma Physics, Chinese Academy of Sciences, Hefei, People's Republic of China
[4] University of Science and Technology of China, Hefei, People's Republic of China
[5] Shihezi University, Shihezi, People's Republic of China
[6] Hengyang Normal University, Hengyang, People's Republic of China
[7] Anhui Province Key Laboratory of Industry Safety and Emergency Technology, Hefei University of Technology, Hefei, People's Republic of China

E-mail: lantao@ustc.edu.cn and zxhui@hfut.edu.cn

**Abstract**

Compact torus injection is considered as a high promising approach to realize central fueling in the future tokamak device. Recently, a compact torus injection system has been developed for the Experimental Advanced Superconducting Tokamak, and the preliminary results have been carried out. In the typical discharges of the early stage, the velocity, electron density and particles number of the CT can reach $56.0\ km/s$, $8.73 \times 10^{20}\ m^{-3}$ and $2.4 \times 10^{18}$ (for helium), respectively. A continuous increase in CT density during acceleration was observed in the experiment, which may be due to the plasma ionized in the formation region may carry part of the neutral gas into the acceleration region, and these neutral gases will be ionized again. In addition, a significant plasma rotation is observed during the formation process which is introduced by the $E \times B$ drift. In this paper, we present the detailed system setup and the preliminary platform test results, hoping to provide some basis for the exploration of the CT technique medium-sized superconducting tokamak device in the future.

Keywords: Compact torus, Central fueling, EAST

## 1. Introduction

In a fusion reactor, most fusion reactions take place in the plasma core region where both the plasma density and temperature are peaked. Pellet enhanced performance (PEP) of tokamak reactors shows that central fueling can substantially increase tokamak confinement, and it also provides an effective method to control the plasma density and pressure profile and has greatly improve burnup efficiency of the main particles [1, 2]. Thus, central fueling is a crucial issue for successful development of magnetic fusion reactors. However, traditional fueling techniques (e.g., gas puffing [3], pellet injection [4], supersonic molecular beam injection [5]) are inefficient and inadequate particularly when central fuelling is required. In order to achieve central fueling in large fusion reactors, alternative fueling technologies must be developed. In comparison, the velocity of compact torus (CT) is typically tens to hundreds of kilometres per second with high density and self-contained magnetized plasma structure [6], which has potential to centrally fuel a reactor-grade tokamak. The basic penetration model is following the equation: $v^2/2 > B_T^2/\mu_0$, where $\rho$ is the mass density of CT, $v$ is the velocity, and $B_T$ is the magnetic field strength of tokamak [4].

The compact torus is an axisymmetric toroidal plasma confinement system which was first discussed as a fusion concept in 1958 by Alfvén [7] and compact torus injection (CTI) using as an emerging technology to centrally fuel a tokamak reactor was originally proposed by Perkins *et al*. [3] and Parks [4] in 1988. Since then, active research has been pursued to validate its feasibility and investigate the interaction between the CT plasma and tokamak plasma. The early experiments on acceleration and focusing of magnetically confined plasma ring were conducted in LLNL Ring Accelerator Experiment (RACE) project in 1988, and the final speed can up to approximately 2500 km/s [8, 9, 10]. The first tokamak fuelling by CT were carried out on Caltech's ENCORE tokamak, and the result shows a large increase in the electron density with a current increase due to helicity injection. But the discharges of the ENCORE tokamak (R = 0.38 m, a = 0.12 m, $B_t = 0.07\ T$) were always disrupted by the CT injection [11]. The first disruption-free CT injection experiment was performed in 1994 on the TdeV tokamak (R = 0.86 m, a = 0.25 m, $B_t = 1.4\ T$) with Compact Torus Fueler (CTF) device [12, 13]. In the following years, various interesting experimental phenomenon induced by the CT injection, such as electrode biasing, turbulent heating and H-mode like discharges establishment, have been discovered [14, 15]. Then in 2006, the first vertical CT injection experiment has been performed in STOR-M tokamak, can achieve deeper penetration, which revealed a prompt increase in the line averaged electron density by more than twofold, 30% reduction in the $H_\alpha$ radiation level, significant suppression of floating potential fluctuations and m = 2 Mirnov oscillations [2, 16, 17]. In addition to tokamak, CT technology has also been tested in other magnetically confined fusion pathways, like field-reversed configuration (FRC) C-2 [18] and Keda Torus eXperiment (KTX) [19]. However, experimental studies on the application of this technology to medium and large tokamak devices have not yet been carried out, and in view of this we developing a CT injection for Experimental Advanced Superconducting Tokamak (EAST) and conducted preliminary bench test experiments.

A schematic diagram of the CT formation and acceleration process is shown in figure 1. First, bias field solenoid provided a bias magnetic flux and the working gas is puffed into the coaxial cylindrical electrode. The gas is breakdown by formation electrode discharge, the discharge current increases rapidly. A current flow $J_r$ through the plasma radially outward inducing a magnetic field $B_\theta$ in the azimuthal direction. The Lorenz force $J_r \times B_\theta$ push the plasma down the electrodes, under sufficient pressure from the $B_\theta$ field, the solenoid field was stretched, forming the CT poloidal field when reconnection occurs [20, 21, 22]. Toroidal magnetic field of the CT is formed by the current loop between the electrodes in the formation region. Within a few microseconds after formation, relaxation yields a long-lived CT configuration approaching the minimum-energy Taylor state [23]. The magnetic field of a CT in a force-free state satisfies the following equation [24]: $\nabla \times B = \lambda B$, where $\lambda$ is a constant. The $J \times B$ force should greater than the magnetic tension of the bias field. The formation current $I_{form}$ and bias magnetic flux $\phi_{bias}$ satisfy the requirement $\lambda = \mu_0 I_{form}/\phi_{bias} > \lambda_c$, a CT will be formed [25]. $\lambda_c$ in relation to the parameters of the device itself following the equation $\lambda_c = (1/r_{in}) * \sqrt{2/ln(r_{out}/r_{in})}$, where $r_{out}$ and $r_{in}$ are the outer and inner radius of formation region. When CT traveling through the acceleration region, the CT is regarded as a sliding short. Acceleration current flows passing through the CT, produce the $J \times B$ force making further acceleration for CT.

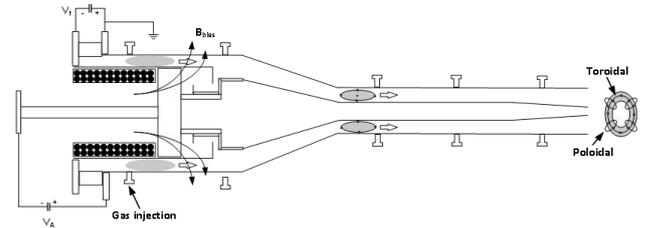

**Figure 1.** Schematic diagram of compact torus formation and acceleration process.

The remainder of this paper is organized as follows: the experimental setup of the CT injector is presented in section 2, typical waveforms and preliminary results of EAST-CTI are given in section 3. Finally, a discussion and summary are provided in section 4.

## 2. Experimental setup

The schematics and photo of the EAST-CTI are shown in figure 2(a)(b). The EAST-CTI consists of formation region, compressor region, acceleration region, gas valve, insulators and solenoid. The power supply of formation and acceleration region is connected to the injection via an external flange. The inner and outer electrodes are separated by insulators. The out shell of EAST-CTI is the only grounding point for all bank systems, and the injection adopts negative high voltage pulse discharge which gives better performance of the CT [26, 27, 28].

The length and diameter of the inner and outer electrodes in formation region are $345\ mm$, $345\ mm$, $209\ mm$ and $255\ mm$, respectively, and they are constructed by 304 stainless-steel. The compressor is $278\ mm$ long, and its radius changes from $125\ mm$ to $50\ mm$ with a compression ratio of 2.5. The process of compression considered to be adiabatic and, therefore, it can make CT with higher density, temperature and magnetic field [11]. The length and diameter of the inner and outer electrodes in acceleration region are $500\ mm$, $500\ mm$, $105\ mm$, and $35\ mm$, respectively, and they are constructed by an oxygen-free copper with better electrical conductivity.

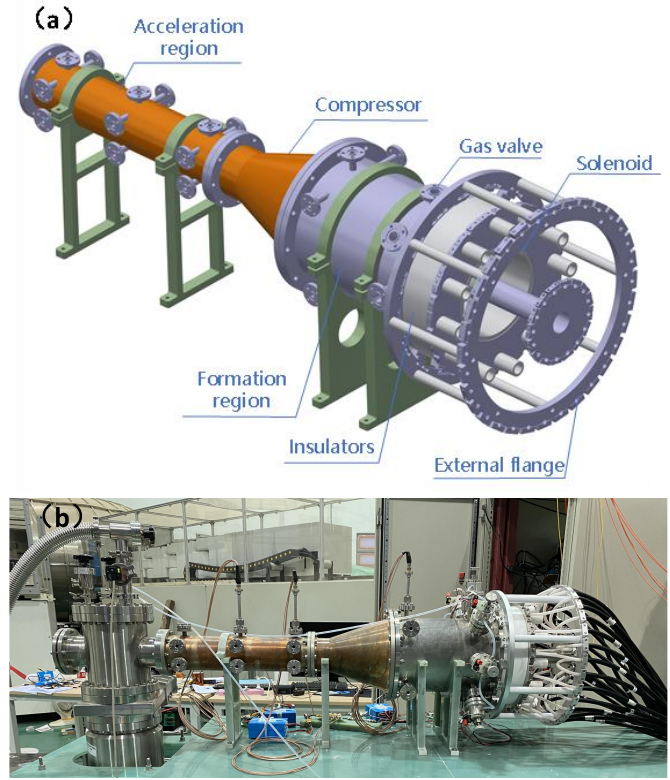

**Figure 2.** (a) Schematic view of EAST-CTI, and (b) picture of the EAST-CTI.

### 2.1 Bias field solenoid

In order to get a CT with spheromak-like plasma rings structure, a solenoid is set insider the inner electrode of the formation region producing a bias magnetic flux. The solenoid is made of copper wire which is wound on a Poly tetra fluoroethylene (PTFE) backbone, the length and thickness of solenoid are $162\ mm$ and $7.6\ mm$, total 810 turns of coil with inductance $80.6\ mH$ and resistance $6.7\ \Omega$. In order to withstand a high voltages applied on the electrode, solenoid surface is covered with polyimide. The solenoid must maintain a magnetic field long enough. In typical CT experiment, the penetration time of the CT does not exceed to tens of microseconds. To ensure that the magnetic field applied to the CT during this period can be approximated as a quasi-steady field.

In figure 3(a), at 2 kV discharge voltage, the solenoid current reached to 137 A. The Full Width at Half Maximum (FWHM) discharge waveform of the solenoid is about $14\ ms$, which is much longer than the main circuit discharge time. Simulation by COMSOL yields the poloidal magnetic flux is $10\ mWb$, poloidal magnetic field at the center of solenoid coil is about 0.5 T.

### 2.2 Fast puff gas valves

Eight gas valves are positioned 45° apart azimuthally to ensure a uniform distribution of pure helium into the formation region, four valves were used in the initial tests. The valve was originally designed by USTC, which has a fast reaction time. The gas valve consist of a round pie coils, an aluminium piston, and a spring, the detailed structure can be found in [29]. Piston is squeezed by spring, and the piston will be pushed upwards by a repulsive force when current passes through the coil. Then the gas enters the chamber. The discharge voltage and current waveforms are shown in figure 3(b), at 2 kV discharge voltage, the current reached to 9000 A and the FWHM of the current waveform is about 90 $\mu s$, basically meets the requirements of the device for particles injection and fast response .

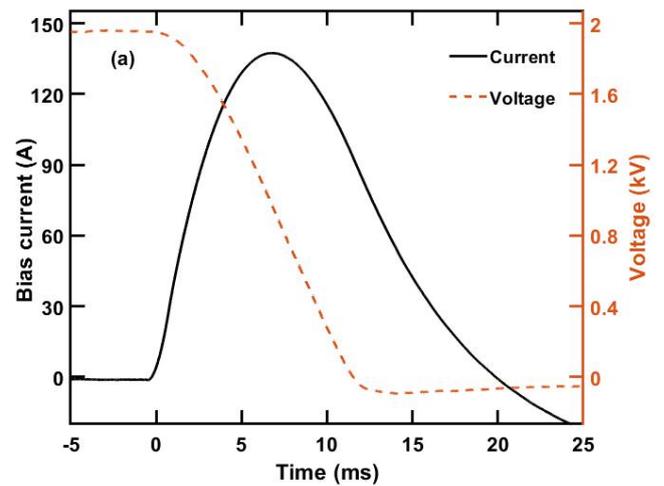

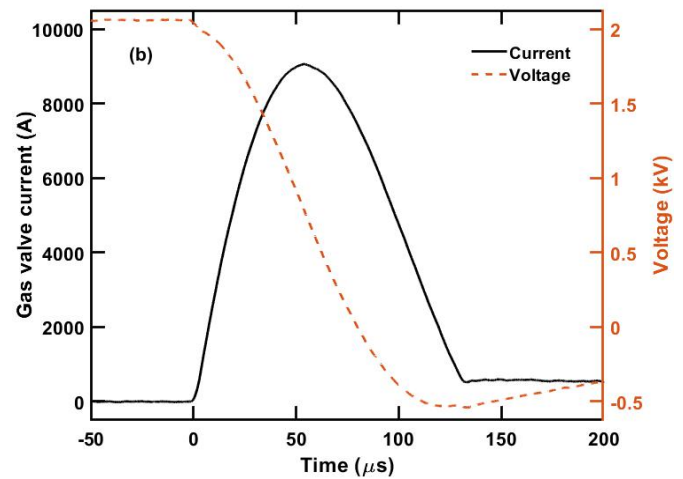

**Figure 3.** (a) Typical discharge of bias field solenoid (charged voltage on bank: 2 $kV$). (b) Typical discharge of gas valve (charged voltage on bank: 2 $kV$).

### 2.3 Power supplies

**Table 1.** Power supplies of EAST-CTI

| Position | Capacitances (µF) | Maximum voltage (kV) |
|---|---|---|
| Formation | 120 | 20 |
| Acceleration | 160 | 20 |
| Bias solenoid | 500 | 2 |
| Gas valve | 250 | 2 |

The power supply system of EAST-CTI is composed of the formation region, acceleration region, bias solenoid and gas valve. The capacitances of formation and acceleration region are $120\ \mu F$ (four $30\ \mu F$ capacitors in parallel) and $160\ \mu F$ (four $40\ \mu F$ capacitors in parallel), which are given in table 1. Maximum discharge voltage of 20 kV for both main circuits, and maximum energy storage are 24 kJ and 32 kJ, respectively. The formation and acceleration regions are connected to the injection by 16 parallel cables, which considerably reduce the inductance of the entire circuit. A $250\ \mu F$ capacitor bank is used to fire these gas valves, and the solenoid is fired using a $500\ \mu F$ capacitor bank. The maximum voltages are $2\ kV$ for both of them.

The bias magnetic flux is provided by solenoid which is installed at formation region, the duration of solenoid magnetic field must be large to penetrate through electrode. The frequency of bias current $f_{bias}$ to penetrate through the electrode is estimated by equation [30]: $f = \rho/\pi\mu\delta^2$, where $\rho$, $\delta$, and $\mu$ are resistivity of the conductor, skin depth, and magnetic permeability, respectively. The thickness of formation electrodes is 3 mm, so the frequency $f_{bias}$ needs to be much slower than 20 kHz. In this solenoid, the frequency of bias current $f_{bias}$ can be estimated from the equation: $f \approx 1/2\pi\sqrt{LC}$, where L and C are inductance and capacitance. The frequency of bias current $f_{bias}$ is approximately 7.9 Hz. So, the frequency of solenoid satisfies for magnetic field to penetrate the electrodes.

*2.4 Diagnostics*

In EAST-CTI, there are various diagnostic are used to monitor discharge waveforms and the motion characteristics of CT plasma. The diagnostics on EAST-CTI are shown in table 2, including Langmuir probes, magnetic probes (MP), Rogowski coils, charge coupled Device (CCD) camera and Impurity spectrometer.

Since the CT is a low temperature, high density plasma, the simplest way to measure CT density is by using Langmuir probe. As shown in figure 4, three Langmuir probes (L1-L3) are located at z=38 cm, $57\ cm$, $79\ cm$, respectively, to monitor the variation of CT density in the acceleration region. In the same position in the axial direction, three magnetic probes (P1-P3) are used to monitor the poloidal magnetic field of the CT. Effective area of MP is about $400\ mm^2$, and the frequency respond can reach up to 1MHz which is sufficient high for CT magnetic measurement. The magnetic fluctuation signal passes through a passive integration ($\tau_{RC} = 50\ \mu s$) to recover the magnetic signal of the CT. The discharge currents of solenoid, gas valve, formation region, acceleration region monitored by Rogowski coils. In order to observe the process of plasma discharge, charge coupled device (CCD) camera (Phantom V710 high-speed camera, resolution 256 × 128, sampling rate 150000 $fps$, 1 $\mu s$ open) is placed at the end of the device for recording. And it is also important to monitor the type of impurities in the CT, the Impurity spectrometer (Wyoptics scientific class spectrometer MAX2000 pro) is placed in the side port of the device.

**Table 2.** Components and number of the diagnostic on EAST-CTI

| Diagnostics | Measurement parameter |
|---|---|
| Magnetic probe | Velocity/Magnetic field strength |
| Langmuir probe | Electron density/ Temperature |
| Rogowski coil | Discharge current |
| CCD camera | Image |
| Impurity spectrometer | Impurity type |

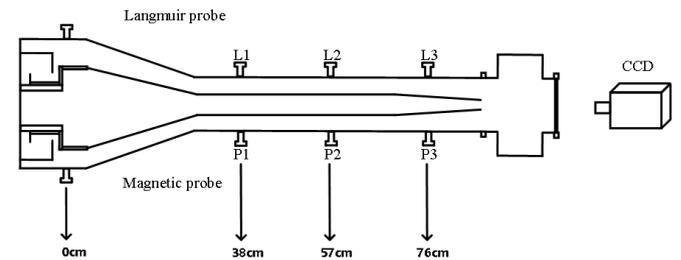

**Figure 4.** Schematic view of the partial diagnosis installed on EAST-CTI

*2.5 Data acquisition and vacuum system*

The data acquisition system consist of six high speed oscilloscope cards PXIe-69834, and its maximum sampling rate is 80 MHz at 16-bit resolution. Each of the six cards has four channels totally with twenty-four high-speed channels. The six oscilloscope cards are mounted on the oscilloscope chassis PXIe-62590 which are connected to the computer via a controller PXIe-63977w. A high vacuum degree is the key to successful discharge, so the electrodes are routinely baked to a temperature of 120 ℃ by using heating tape. A turbo molecular pump (FF-200) is used, and its pumping speed is 1200 L/s. The ultimate base pressure is about $1 \times 10^{-5} Pa$.

**3. preliminary results**

*3.1 Typical waveforms*

The precise control of time is important for CT pulse discharge. For this purpose, a set of Field Programmable Gate Array (FPGA) based high-precision timing control module had developed. A typical discharge sequence is shown in figure 4. In the actual discharge, the solenoid discharge time is set at $-7000\ \mu s$ just as shown in figure 5(a). At the charging voltage $0.8\ kV$, bias solenoid current is about $40\ A$, the poloidal magnetic field is about $0.17\ T$ and the bias magnetic flux is about $3\ mWb$. At normal temperature, the adiabatic expansion velocity of helium in formation region is approximately 1.24 km/s [31], and the distance from the gas valve to the end of the formation region is $0.25\ m$. Therefore, in order to ensure the homogeneous diffusion of the gas and the gas does not spread outside the formation region. The gas valve is opened earlier 200 μs than the discharge of the formation region, so the discharge moment of the gas valve is chosen at $-200\ \mu s$. In figure 5(b), typical gas valve current is approximately $3700\ A$ with the charging voltage $V_{gas} = 1\ kV$. The moment of formation region discharge is considered to be the zero moment. As shown in figure 5(c)(d), the formation and acceleration discharge currents are $144\ kA$ and $87\ kA$, with discharge voltage at $6\ kV$ and $5\ kV$, respectively.

The value of $\lambda$ is $60\ m^{-1}$ in this discharge, and a necessary value of $\lambda_c$ is derived from the size of the injection, which is $30.3\ m^{-1}$. Hence, the parameter satisfied the need to form and accelerate a CT.

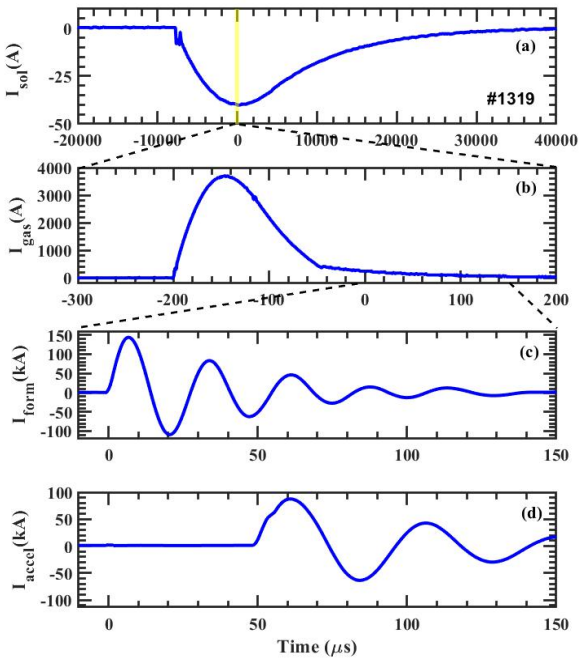

**Figure 5.** Typical discharge waveforms. (a) Solenoid bank current ($V_{sol} = 0.8\ kV$). (b) Gas valve bank current ($V_{gas} = 1\ kV$). (c) Formation bank current ($V_{form} = 6\ kV$). (d) Acceleration bank current ($V_{accel} = 5\ kV$).

### 3.2 Experimental results

Figure 6 (a-d) shows the relationship between acceleration region current and CT density at different locations in acceleration region. The maximum electron densities at $Z = 38, 57, 76\ cm$, are $3.91 \times 10^{20} m^{-3}$, $6.16 \times 10^{20} m^{-3}$ and $8.73 \times 10^{20} m^{-3}$, respectively. The velocity of $v_{1,2}$, $v_{2,3}$, which represent the velocity between $Z = 38 - 57 cm$ and $Z = 57 - 76 cm$, are equal to $39.3\ km/s$ and $56.0\ km/s$. The velocity growth rate of CT between these two distances about 42.5%. At the same time, experimental observation the density of CT is increasing from $3.91 \times 10^{20} m^{-3}$ at $38\ cm$ to $8.73 \times 10^{20} m^{-3}$ at $76\ cm$. This is maybe due to the neutral gases which carried from formation region ionised again, and it is beneficial for local plasma density increase.

The poloidal magnetic field $B_p$ represent CT has been formed. Three magnetic probes located in the acceleration region are used to measure the magnetic field. Figure 6(e-g) shows the changes of boundary magnetic field at same axially position. The negative peak of $B_{p1}$, $B_{p2}$, and $B_{p3}$ signal are continuous decay, which means that the CT is also decaying during transmission.

To estimate the totally number of particles, we use the equation from $\int n_e dV = \int S_{CT} v_{CT} n_e dt$, where V is the volume of CT, $S_{CT}$ is the particle flow cross-sectional area, which is about $0.007\ m^2$. The total particle number, mass and length of the CT is $N_p \approx \int S_{CT} V_{CT} n_e(t)\ dt = 2.4 \times 10^{18}$, $m_{CT} \approx 16.1\ \mu g$ and $L \approx v_{CT} \times t_{fwhm} = 0.17\ m$. In general, we assume that the density of CT is uniformly distributed. The injected momentum, total directional kinetic energy and penetration pressure are $P_{CT} \approx m_i N_p v_{ct} = 9 \times 10^{-4}\ N \cdot s$, $E_k \approx \frac{1}{2} m_i N_p v_{CT}^2 = 25\ J$ and $P_k \approx \frac{1}{2} m_i n_e v_{CT}^2 = 1.6 \times 10^3\ Pa$, respectively. The corresponding maximum penetrating magnetic field is $B_{max} = \sqrt{2\mu_0 P_k} = 0.06\ T$.

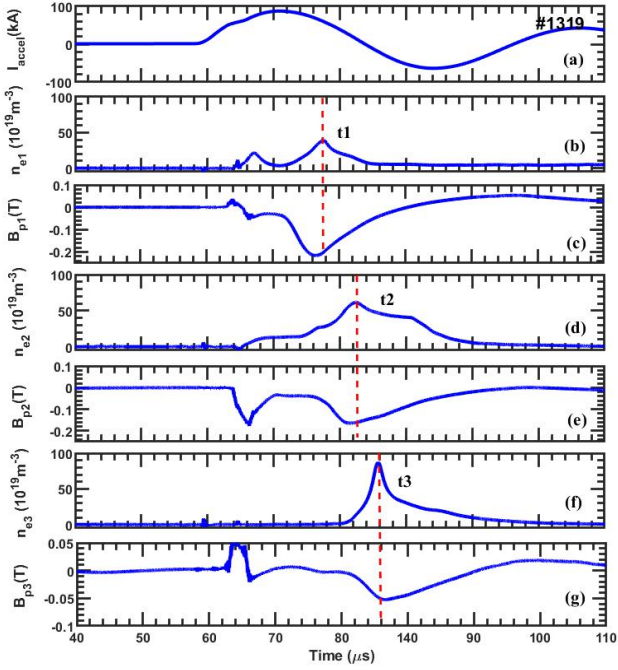

**Figure 6.** The experiment results of CT: (a) acceleration current waveform, (b-d) electron density and (e-g) magnetic field strength.

　　In order to obtain the process of formation and evolution of CT during the discharge, the image is captured by high-speed camera with an interval of $6.67\ \mu s$. The begin time of formation region discharge is chosen as the zero-time trigger of the camera. In figure 7(a-d), the breakdown of the neutral gas is observed at $t = 6.67\ \mu s$, and four channels of current are formed in the local area. Subsequently, the four plasma clusters spread rapidly in radial direction and sweep the neutral particles. At $t = 13.33\ \mu s$ two bright clusters of light are observed on images and rotating with the velocity near the inner central electrode. The bright cluster underwent a significant rotation at $t = 20\ \mu s$ compared to $t = 13.33\ \mu s$. Using the equation $v_\theta = 2\pi r_{in}\theta/\pi T = 16\ km/s$, where θ is the angle of rotation and T is the interval between these times. The possible causes of this rapid rotation are bright plasma cluster subject to a force $E_r \times B_z$ drift motion in the toroidal direction (formation electrodes voltage is 1.2 kV, and the bias field is 0.17 T) [32]. And at $26.67\ \mu s$ there are no bright spots on the plasma cluster, which means the isomer cluster is uniformly mixed overall. The rapid rotation of the plasma accelerates the homogeneous mixing of gases, which is beneficial for CT formation and acceleration.

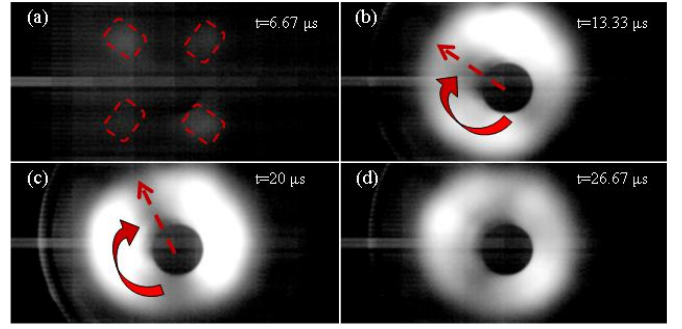

**Figure 7.** The images of the CT during discharge at different solenoid bias field (a1-d1).

　　Impurities have a significant impact on the quality and speed of the CT, therefore, the type of impurities produced during discharge must be measured. In Figure 8, the dominant wavelengths in the range $900 - 1100\ mm$ were measured by an impurity spectrometer and calibrated to obtain the main compositions of $C$, $Cu$ and $Fe$. The attenuation of the CT magnetic field is mainly caused by the impurities, and also greatly affects the lifetime of the CT. The usual solution is to coat the walls of the device with tungsten of chromium, which are not easily ionized.

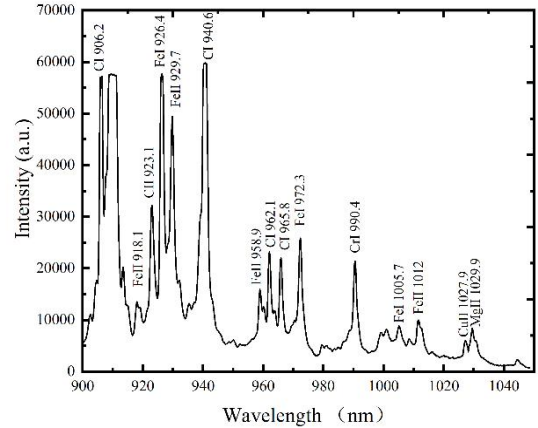

FIG.8. Measurement result of the dominant wavelength by using the impurity spectrometer.

## 4. Summary and Discussion

　　This paper presents a newly designed compact torus injector (EAST-CTI) for central fueling of EAST, and the first laboratory measurement are presented. Experimental results show that plasma ionized in the formation region bring part of the neutral gas into the acceleration region, and the neutral gas is ionized again, which is beneficial for increase local plasma density. In summary, the newly developed EAST-CTI can produce a compact torus, and the velocity, electron density, particle inventory, mass, length, injected momentum, total directional kinetic energy and penetration pressure of CT are $56.0\ km/s$, $8.73 \times 10^{20} m^{-3}$, $2.4 \times 10^{18}$, $16.1\ \mu g$, $0.17\ m$, $9 \times 10^{-4}\ N \cdot s$, $25\ J$ and $1.6 \times$ $10^3\ Pa$, respectively. In future, and the experimental procedure will be optimized including increase the discharge voltage to obtain higher parameters.

**Acknowledgements**

This work is supported by the National Key Research and Development Program of China under Grant Nos. 2017YFE0300500, 2017YFE0300501, and the Institute of Energy, Hefei Comprehensive National Science Center under Grant Nos. 19KZS205, 21KZS202, and the University Synergy Innovation Program of Anhui Province under Grant Nos. GXXT-2021-014, GXXT-2021-029. This work is also supported by the National Natural Science Foundation of China under Grant No. 11905143, and the Users with Excellence Program of Hefei Science Center CAS 2020HSC-UE008, and the Fundamental Research Funds for the Central Universities of China under Grant No. PA2021GDSK0073.